\documentclass[aps,10pt,pra,twocolumn,amsmath,amssymb,superscriptaddress,groupedaddress]{revtex4-1}

\usepackage[english]{babel} 
\usepackage[utf8]{inputenc}
\usepackage[T1]{fontenc}

\usepackage{bbm}
\usepackage{verbatim}
\usepackage{graphicx}
\usepackage{times}
\usepackage{epsfig}
\usepackage{graphicx}
\usepackage{bm}
\usepackage{txfonts}
\usepackage{dsfont}
\usepackage{color}
\usepackage{braket}

\newcommand\at[2]{\left.#1\right|_{#2}}

\usepackage{hyperref}
\hypersetup{
colorlinks=true,
linkcolor=blue,          % color of internal links
citecolor=blue,          % color of links to bibliography
filecolor=magenta,       % color of file links
urlcolor=blue
}

\begin{document}

\selectlanguage{english}

\title{Optomechanical Self-Oscillations in an Anharmonic Potential:\\ Engineering a Nonclassical Steady State}

\author{Manuel Grimm}
\author{Christoph Bruder}
\author{Niels Lörch}
\affiliation{Department of Physics, University of Basel, Klingelbergstrasse 82, CH-4056 Basel, Switzerland}

\date{\today}

\begin{abstract}
  We study self-oscillations of an optomechanical system, where
  coherent mechanical oscillations are induced by a driven optical or
  microwave cavity, for the case of an
  anharmonic mechanical oscillator potential. A semiclassical
  analytical model is developed
  to characterize the limit cycle for large mechanical
  amplitudes corresponding to a weak nonlinearity. As a result, we
  predict conditions to achieve subpoissonian phonon statistics in the
  steady state, indicating classically forbidden behavior. We compare
  with numerical simulations and find very good agreement. Our model
  is quite general and can be applied to other physical systems such
  as trapped ions or superconducting circuits.
\end{abstract}

\maketitle

\section{Introduction}

The state of individual physical systems is determined by the
interaction to their environment. Most natural environments randomly
couple the system to many degrees of freedom and bring about classical states \cite{Joos2003, Zurek2003c}. 
But artificial environments can be specifically engineered \cite{Poyatos1996}, typically by strongly coupling the system to a small set of well-controlled degrees of freedom, for the purpose of reaching a particular steady state that may have nonclassical features. Such states are a crucial resource for quantum information processing \cite{Nielsen2010, Mari2012b}, furthermore they are of fundamental interest for testing quantum mechanics in previously unexplored regimes \cite{Bassi2013}.

Quantum reservoir engineering has been used to demonstrate
nonclassical steady states on various platforms such as atomic clouds
\cite{Krauter2011}, superconducting qubits \cite{Murch2012,
  Shankar2013}, and trapped ions \cite{Kienzler2015}. In the context
of optomechanics, driving the optical cavity on both sidebands can
lead to highly nonclassical states. Steady-state mechanical squeezed
states have been proposed \cite{Kronwald2013} and realized
\cite{Wollman2015, Pirkkalainen2015, Lecocq2015a}
by driving
dominantly on the red sideband. For dominant blue sideband driving, stabilization of mechanical Fock states has been proposed \cite{Rips2012a}, requiring in addition a strongly intrinsic mechanical anharmonicity, which has not been realized in mechanical oscillators.

For weaker anharmonicity, such a setup with dominant driving on the blue sideband leads to coherent excitation of mechanical self-oscillations and therefore laser-like mechanical states, which we investigate in this article.
For the case in which the intrinsic anharmonicity is the system's dominant nonlinearity,
we derive a semiclassical analytical description to describe the system dynamics in terms of the amplitude. The description is valid for large mechanical amplitudes, where we compare to numerical simulations and find excellent agreement.

For such a setup we derive conditions on the system parameters for the steady states to show number squeezing, which is characterized by subpoissonian number statistics. 
This nonclassical feature is well-studied in the photon statistics of
lasers and can be achieved e.g. by pumping the cavity with an ordered
sequence of
separated flying atoms \cite{Sokolov1984} or coupling to one-and-the-same fixed atom \cite{McKeever2003}. The subpoissonian statistics in these system is in contrast to ordinary lasers, where the random pumping via a large number of atoms results in fully classical coherent or even superpoissonian states. 

In the context of optomechanical self-oscillations
\cite{Marquardt2006}, recently several proposals have been made to
achieve the analogous phenomenon for phonons,  i.e. subpoissonian
statistics for a phonon laser \cite{Rodrigues2010, Qian2012,
  Armour2012b, Nation2013, Lorch2014a, Lorch2015}. All of these
proposals rely on the nonlinearity of the optomechanical
interaction. In contrast, we consider a linearized optomechanical
interaction and use the intrinsic nonlinearity of the mechanical oscillator to achieve subpoissonian statistics.
 
While we will employ optomechanical terminology throughout this
article, the underlying model is quite general and can be applied to
other implementations. For example, phonon lasing has been demonstrated with trapped ions \cite{Vahala2009, Knunz2010a} and
ion potentials can be engineered to have a large nonlinearity, so that
the dynamics will be similar to the discussion in this paper. A further implementation could be done with superconducting circuits, which can have large effective Kerr nonlinearities.

\begin{figure}[t]
\centering 
\includegraphics[width=0.4\textwidth]{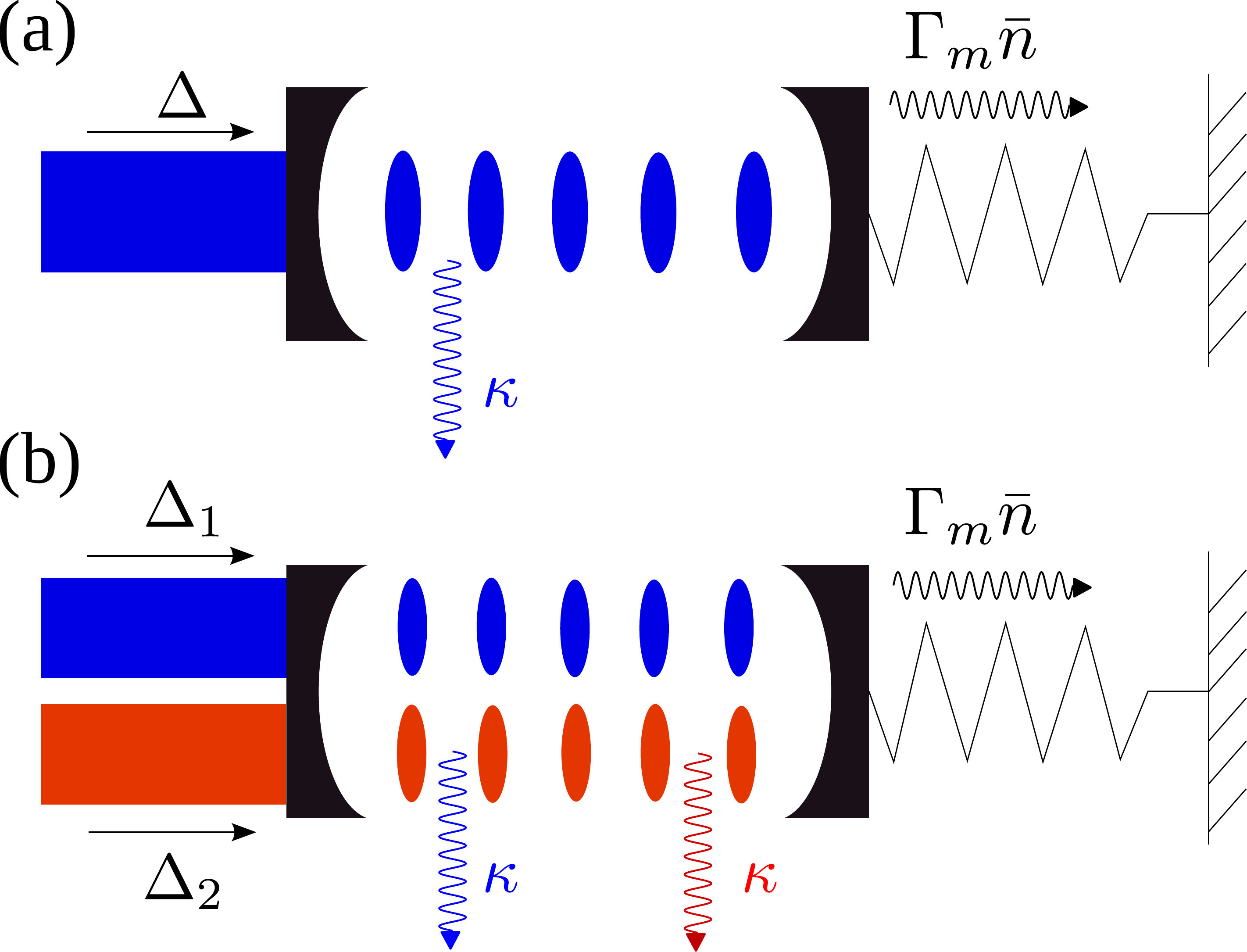}
\caption{
\label{SetupFig}
Illustration of the studied systems. (a) A standard optomechanical
system, where an optical cavity mode with decay rate $\kappa$ is
dispersively coupled to a mechanical mode of resonance frequency
$\omega_m$, decay rate
$\Gamma_m$ and bath occupation $\bar n$.
 The cavity is driven by a laser on the blue sideband $\Delta \approx \omega_m$ to excite coherent mechanical oscillations. (b) Another laser tone with detuning $\Delta_2=-\omega_m$ is added on the red side to reduce the effective temperature of the mechanical bath. 
 } 
\label{fig: kerr schematic setup}
\end{figure}

This article is structured as follows:
In the following Section \ref{OneCav} we develop an analytical description for the system. For simplicity this is done for the case illustrated in Fig. \ref{SetupFig} (a), where only one laser drives the cavity.   
In Section \ref{sec: Two independent Cavities} we then generalize to the two-laser case depicted in Fig. \ref{SetupFig} (b). Finally in Section \ref{Disc} we discuss these results and compare to numerical solutions of the quantum master equation.

\section{Analytical description}

\label{OneCav}

In this section we derive the main results for a system that is driven by only one laser and generalize later in Section \ref{sec: Two independent Cavities} to the case of two lasers.

\subsection{Model}

\label{model}

We consider a bosonic mode $b$ of a mechanical oscillator with intrinsic Kerr anharmonicity that may be described by the Hamiltonian
$H_m= \omega_m b^\dagger b + K(b^\dagger b)^2$
coupled to a driven optical cavity mode $a$ with Hamiltonian
$H_c=\omega_c a^\dagger a +\Omega \left(a e^{i\omega_L t } + a^\dagger e^{-i\omega_L t} \right)$.
Here $K$ is the Kerr anharmonicity parameter and $\omega_m$, $\omega_c$ and $\omega_L$ are the frequencies of the mechanical and the optical mode, as well as the optical drive of strength $\Omega$. The operators $a^\dagger,a $ and $b^\dagger,b$ denote the creation and annihilation operators for the optical cavity and the mechanical oscillator. The Kerr anharmonicity approximates an anharmonic Duffing term $\propto D (b+b^\dagger)^4$ in the potential, the validity of this rotating wave approximation is discussed in Section \ref{Impl}.

The optomechanical interaction is described by $H_\mathrm{int} = -g_0 a^\dagger a (b+b^\dagger)$ with single-photon coupling $g_0$. Defining the detuning $\Delta=\omega_L-\omega_c$, we switch to a rotating frame for the laser to obtain a time-independent Hamiltonian $H_c=-\Delta a^\dagger a +\Omega (a + a^\dagger)$, while $H_m$ and $H_\mathrm{int}$ are unchanged.
 In the limit of small $g_0$ and large number of photons $n_c$ in the cavity, we further simplify the Hamiltonian
and linearize \cite{Aspelmeyer2014} the interaction to
$H_I= - g(a+a^\dagger) (b+b^\dagger)$, where $g=g_0\sqrt{n_c}$ is the linearized coupling, so that in total $H=H_m+H_c+H_{I}$ is
\begin{equation}
H=\omega_m b^\dagger b + K(b^\dagger b)^2-\Delta a^\dagger a +\Omega \left(a + a^\dagger \right) - g(a+a^\dagger) (b+b^\dagger)\:.
\end{equation}
Note that we already neglected a constant force $\propto g_0 \langle
a^\dagger a \rangle$ which results in a small shift of the mean
position of the oscillator.

The incoherent coupling of the system to its environment can be modeled \cite{Aspelmeyer2014} with the Lindblad operators
\begin{align}
\label{dissipators}
\mathcal{L}_m \rho = &-\Gamma_m(n_{\mathrm{th}}+1) (b^\dagger b \rho + \rho b^\dagger b-  2 b \rho b^\dagger) \nonumber \\
&- \Gamma_m n_{\mathrm{th}} (b b^\dagger \rho + \rho b b^\dagger  - 2b^\dagger \rho b)\:, \nonumber \\
\mathcal{L}_c \rho =& -\kappa (a^\dagger a \rho + \rho a^\dagger a- 2 a \rho a^\dagger)\:,
\end{align}
where $\kappa$ and
$\Gamma$ are the amplitude decay rates of the
cavity and the mechanical oscillator. We assumed here a
zero-temperature bath for the optical cavity and a thermal occupation
$n_{\mathrm{th}}$ of the mechanical bath.
Including these Lindblad operators, the full quantum master equation for this system reads
\begin{equation}
\dot \rho = -i[H, \rho] + \mathcal{L}_m \rho + \mathcal{L}_c \rho\:. 
\label{meq1}
\end{equation}

To obtain a semiclassical description we transform the quantum master equation
\eqref{meq1} into a partial differential equation for the Wigner distribution $W(\beta,\beta^*)$ using the translation rules \cite{Gardiner2004b}
$
b \rho \rightarrow (\beta + q \partial_{\beta^*})W$, 
$b^\dagger \rho \rightarrow (\beta^*-p\partial_\beta)W$
and their complex conjugates to obtain
\begin{equation}
\label{eq: dt W}
\begin{split}
\partial_t W =& \big[-i{\Delta}\partial_\alpha \alpha +i \omega_m \partial_\beta \beta -ig \big( \partial_\alpha(\beta + \beta^*)+\partial_\beta(\alpha +\alpha^*)   \big)\\
&+\Gamma_m (\partial_\beta \beta +\tfrac 12(2\bar{n}+1)\partial_\beta \partial_{\beta^*})+ \kappa(\partial_\alpha \alpha + \tfrac 12 \partial_\alpha \partial_{\alpha^*} )\\
&+iK\partial_\beta(2|\beta|^2\beta -\beta - \tfrac{1}{8}  \partial_\beta^2\partial_{\beta^*} \beta) \big]W +\mathrm{h.c.} \:.
\end{split}
\end{equation}
Assuming large mechanical amplitudes we neglect third-order derivatives in a truncated Kramers-Moyal expansion \cite{Carmichael99} so that Eq.~(\ref{eq: dt W}) becomes a Fokker-Planck equation. Its corresponding Langevin equations are
\begin{align}
\label{eq: alpha_dot}
&\dot{\alpha}=(i {\Delta}-\kappa)\alpha +ig (\beta+\beta^*) + \eta_\alpha, \\
&\dot{\beta}=\big(-i\omega_m-iK(2|\beta|^2-1)-\Gamma_m\big)\beta + ig (\alpha +\alpha^*) + \eta_\beta\:,
\label{eq: eqs of motion}
\end{align}
where $\eta_{\alpha}, \eta_{\beta}$ are zero-mean complex white noise processes with the correlators 
$\braket{\eta_\alpha(t) \eta_{\alpha^*}(t')}=\kappa \delta(t-t')$, 
$\braket{\eta_\beta(t)
  \eta_{\beta^*}(t')}=\Gamma_m(2\bar{n}+1)\delta(t-t')$,
and 
$\braket{\eta_i(t) \eta_i(t')}=\braket{\eta_i(t) \eta_{j^*}(t')}=0$
for $i,j \in \{\alpha, \beta\}$ and $i \neq j$.

\subsection{Adiabatic Elimination of the Cavity}
To eliminate the optical amplitude $\alpha$,
we assume the cavity decay rate to be much greater than the interaction strength and the mechanical damping, i.e. $\kappa \gg g, \Gamma_m$, and furthermore we assume that the mechanical frequency is much larger than the interaction strength $\omega_m \gg g$. These are realistic assumptions that can be achieved in typical optomechanical experiments. For the mechanical amplitude we choose the ansatz 
$
\beta=B e^{-i\phi}e^{-i\omega_m(B)t}
$, with 
\begin{equation}
\omega_m(B)=\omega_m+2KB^2-K\:,
\end{equation}
where $\phi(t)$ and $B(t)$ are real-valued numbers describing the phase and
amplitude of the oscillator. According to our assumptions they are
slowly varying on the time scale of
$\kappa^{-1}$. In contrast to the
otherwise quite analogous treatment of optomechanical limit
cycles given in \cite{Marquardt2006,Rodrigues2010, Armour2012b}, we
have to choose here an amplitude-dependent frequency $\omega_m(B)$
because of the factor $-i K(2|\beta|^2-1)$ in the equation of motion~(\ref{eq: eqs of motion}). Defining the Fourier transform  as
$
\mathcal{F}[f(t)] = \int \mathrm{d}t e^{-i\omega t} f(t), \quad \mathcal{F}^{-1}[g(\omega)] = \frac{1}{2\pi} \int \mathrm{d}\omega e^{i\omega t} g(\omega),
$
we can solve Eq.~\eqref{eq: alpha_dot} for $\alpha=\braket \alpha + \delta \alpha$ by adiabatic elimination to obtain
\begin{align}
\label{eq: alpha}
&\braket \alpha(t) =  ig \left( \frac{\beta(t)}{-i\omega_m(B)-i\Delta+\kappa}  + \frac{\beta^*(t)}{i\omega_m(B)-i\Delta+\kappa}\right),\\
&\delta \alpha(t) = \mathcal{F}^{-1} \left[ \frac{\eta_\alpha(\omega)}{i\omega-i\Delta+\kappa} \right].
\end{align}
Inserting Eq.~(\ref{eq: alpha}) into Eq.~(\ref{eq: eqs of motion}) but neglecting the terms $\sim \beta^*$ in a rotating-wave
approximation \cite{Walls2008}, since they will rotate at a frequency
$2\omega_m(B)$ with respect to $\beta$, we find the equation of motion
\begin{equation}
\label{eq: eq motion beta}
\begin{split}
\dot \beta =& -(i\omega_m(B) + i\delta\omega +\Gamma_m + \Gamma_\mathrm{opt}) \beta + \eta_\beta 
+ ig\left(\delta \alpha +\delta \alpha^*\right)\:.
\end{split}
\end{equation}
Here, we defined the optically induced damping and frequency shift
\begin{align}
\label{eq: Gamma_opt}
&\Gamma_{\mathrm{opt}}(\Delta,B)=g^2 \left(\frac{\kappa}{({\Delta}+\omega_m(B))^2+\kappa^2} -\frac{\kappa}{({\Delta}-\omega_m(B))^2+\kappa^2}\right), \\
&\delta \omega (\Delta,B)= g^2 \left( \frac{\omega_m(B)+ \Delta }{( \Delta +\omega_m(B))^2+\kappa^2} + \frac{\Delta -\omega_m(B)}{( \Delta -\omega_m(B))^2+\kappa^2} \right).
\end{align}
These results are analogous to the standard linearized optomechanical Hamiltonian \cite{Aspelmeyer2014}, but with amplitude-dependent frequency. 
Next we switch to polar coordinates and focus on the equation of motion for the amplitude
\begin{align}
\label{eq: eq motion B}
&\dot{B}= -(\Gamma_m + \Gamma_{\mathrm{opt}})B + \eta_T^-, \\
&\eta_T^- = \tfrac 12 \eta_\beta e^{i\varphi}e^{i\omega_m(B)t} 
-{g} 
 \sin\left( {\varphi}+{\omega_m(B)t}  \right) \delta \alpha   +\mathrm{h.c.},
\end{align} 
where $B=|\beta|$ and $\eta_T^-$ refers to the noise in radial direction.

\begin{figure}[t]
\includegraphics[width=0.25\textwidth]{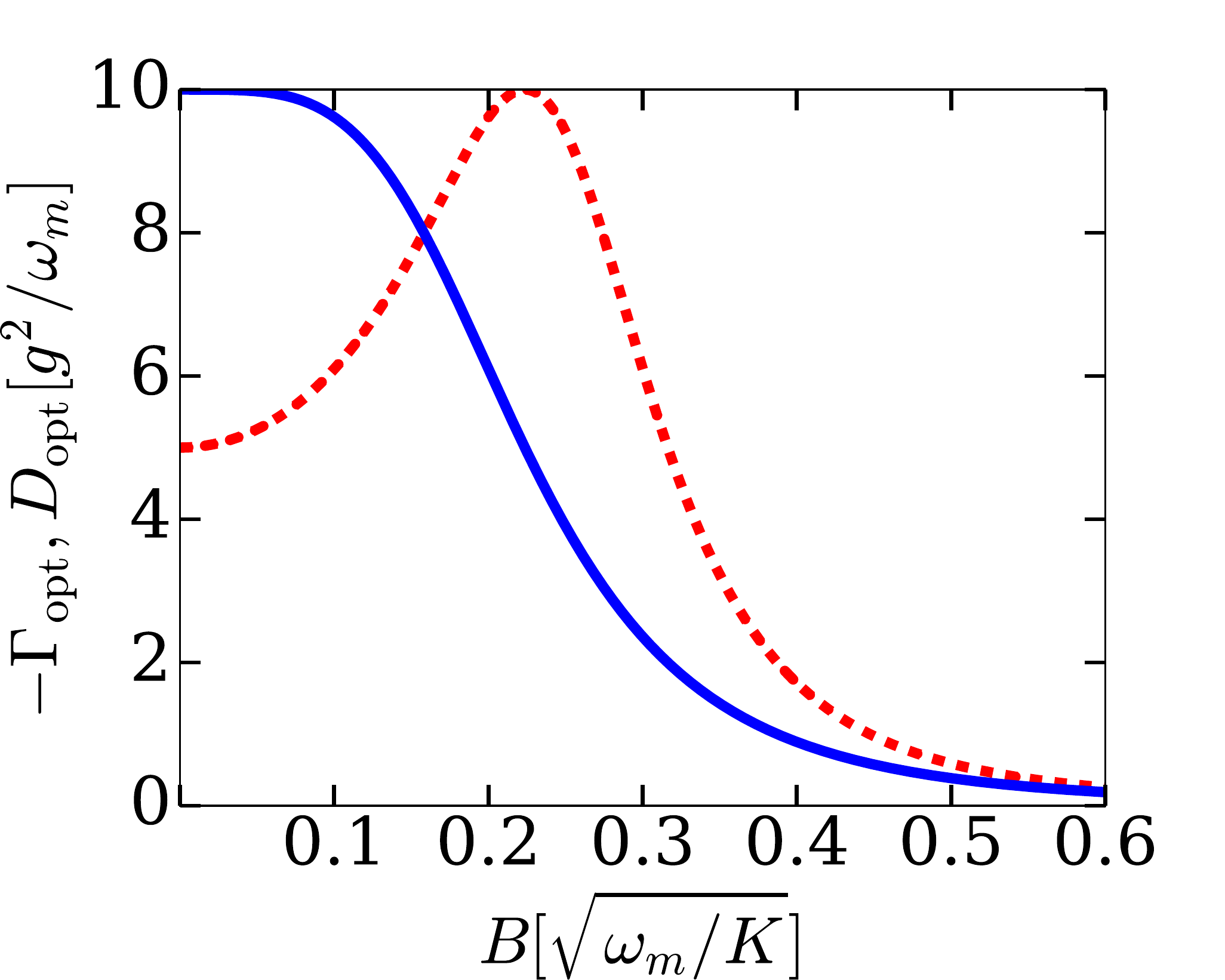}
\caption{
\label{OneLaserDD}
Optically induced damping $\Gamma_{\mathrm{opt}}$ and diffusion $D_{\mathrm{opt}}$ as a function of the amplitude $B$ for the setup with one cavity. The damping is  equal to the negative diffusion, as we neglected the off-resonant terms in Eqs.~\eqref{eq: Gamma_opt},~\eqref{DOPT}. The parameters in this plot are $\kappa/\omega_m =0.1$,  $K/\omega_m \to 0$, $\Delta=\omega_m$ (blue solid line),  $\Delta=1.1 \omega_m$ (red dashed line).
}
\end{figure}

Following \cite{Balanov2008} we evaluate the diffusion constant 
$
D_B = 2\int \mathrm{d}\tau \braket{\eta_{T}^-(t), \eta_{T}^-(t+\tau)} 
$
to convert Eq.~(\ref{eq: eq motion B}) into an effective Langevin equation 
$
\dot{B}=-(\Gamma_m +\Gamma_{\mathrm{opt}})B+ \sqrt{D_B} \eta_B, 
$
where $\eta_B$ is a Gaussian white-noise process. Since this equation is independent of the phase $\varphi$ we can also write down a Fokker-Planck equation for the amplitude probability distribution $W_B$
\begin{align}
\partial_t W(B) = -\partial_B  A_B  W(B) + \frac 12 \partial_B^2 D_B W(B) \label{eq: W_B}
\end{align}
with drift $A_B=-(\Gamma_m + \Gamma_{\mathrm{opt}})B$ for the radial coordinate.  In total we have $D_B = \frac{1}{2}(D_m + D_\mathrm{opt})$, where $D_m=\Gamma_m(2\bar{n}+1)$ refers to the intrinsic mechanical part. After integration we find the optically induced part of the amplitude diffusion
\begin{equation}
\label{DOPT}
D_\mathrm{opt}(\Delta,B) = g^2 \left( \frac{\kappa}{( \Delta +\omega_m(B))^2+\kappa^2}+\frac{\kappa}{( \Delta -\omega_m(B))^2+\kappa^2} \right),
\end{equation}
again deviating from the well-known results in linearized optomechanics only by the amplitude dependence of $\omega_m(B)$. Both optically induced damping \eqref{eq: Gamma_opt} and diffusion \eqref{DOPT} are given by the same Lorentzian as illustrated in Fig.~\ref{OneLaserDD}.

\subsection{Steady-State Solution and Fano Factor}
We have derived an effective equation of motion in the form of a Fokker-Planck equation for the amplitude $B$. We will now calculate its steady-state solution. 
The analytical solution of the Fokker-Planck equation~(\ref{eq: W_B}) is given by \cite{Carmichael99}
\begin{equation}
\label{eq: steady state}
W_{B} = \mathcal{N} \frac{1}{D_B} \mathrm{exp}\left(2 \int \limits_0^B \frac{A_{ B'}}{D_{{B'}}} \mathrm d B'\right),
\end{equation}
where $\mathcal{N}$ is a normalization constant. 
Rather than calculating the full solution, it is more instructive
to analyze the solution after the following approximations. The center $B_0$
of the amplitude distribution obeys the fourth-order equation
\begin{equation}
A_B(B_0) = -\left[\Gamma_m + \Gamma_\mathrm{opt}(B_0)\right]B_0 = 0\:,
\end{equation}
see the definition of the optical damping in Eq.~(\ref{eq:
  Gamma_opt}). This can be simplified by assuming  
$(\omega_m(B_0)-\Delta)^2 \ll (\omega_m(B_0) +\Delta)^2$ and approximating
$
\Gamma_\mathrm{opt} \approx - \frac{g^2 \kappa}{(\omega_m(B)-\Delta)^2 + \kappa^2}
$
by dropping the non-resonant term.
With this simplification the average amplitude in the steady state reads
\begin{equation}
\label{eq: B_0}
\begin{split}
B_0
=\sqrt{\frac{1}{2K}\left( \Delta-\omega_m +K+ \kappa \sqrt{C-1} \right)}\:,
\end{split}
\end{equation}
where $C \equiv {g^2}/{\kappa \Gamma_m}$ is the cooperativity and
we used the conditions $C \geq 1$ and $\Delta > -\omega_m +K+ \kappa \sqrt{C-1}$.
Equation~\eqref{eq: B_0} is a good approximation for the parameter
regime considered here ($\omega_m \gg \kappa \gg g,K \gg \Gamma_m$),
as long as the mechanical damping
$\Gamma_m$ is not too small.
The amplitude $B_0$ scales inversely with $K$, i.e. $B_0$ is larger for small nonlinearities as was expected. 
Note that for very large detunings this expression is not valid as the limit cycle will not start.

Since we expect only small fluctuations around the mean of the amplitude distribution, we linearize the drift around $B_0$. Using $\Gamma_m + \Gamma_\mathrm{opt}(B_0)=0$ we find
\begin{equation}
A_B(B) \approx A_B(B_0) + \at{\frac{\mathrm{d}A_B}{\mathrm{d}B}}{B=B_0}\delta B  = -\Gamma_L \delta B\:,
\end{equation}
where we defined the amplitude fluctuation $\delta B = B-B_0$ and the linearized damping
$
\Gamma_L =  B_0 \at{\frac{\mathrm{d}\Gamma_\mathrm{opt}}{\mathrm{d}B} }{B=B_0}
$.
The steady-state solution Eq.~(\ref{eq: steady state}) is then the Gaussian distribution \cite{Rodrigues2010}
$
W_{B} \sim \mathrm{exp}\left( -\frac{1}{2} \frac{(\delta B)^2}{\sigma^2} \right)
$
 with mean $B_0$ and variance 
 \begin{align}
 \label{eq: W_{B,SS}}
\sigma^2 = \frac{D(B_0)}{2 \Gamma_L}\:.
\end{align}
Based on this approximate solution
we derive conditions under which the oscillator shows number squeezing
and is therefore in a nonclassical steady state. This can be
quantified by the Fano factor
\begin{align}
  F=\frac{\braket{n^2}-\braket{n}^2}{\braket{n}}\;,
\end{align}  
the variance divided by the mean of the phonon number $n=b^\dagger b$.
A Fano factor smaller than $1$ implies subpoissonian phonon statistics, i.e. number-state squeezing. 

To derive the mechanical Fano factor, we make use of the Wigner function to calculate expectation values of symmetrically ordered products of annihilation and creation operators $b, b^\dagger$, e.g. 
$
\braket{B^2}_W = \braket{n} +  \frac{1}{2}$
and
$
\braket{B^4}_W = \braket{n^2}+ \braket{n}+\frac{1}{2}
$,
where the expectation values of the operators $n$, $n^2$ are taken with respect to the steady-state density matrix and the expectation values of $B, B^2$ are with respect to the corresponding Wigner function. 
In the large-amplitude limit the Fano factor can then be rewritten in
terms of the amplitude $B$ as $F \approx 4\sigma^2$.

\label{sec: Fano Factor}

For blue detuning we drop the non-resonant term in $D_\mathrm{opt}(B_0)$ and approximate
$
D_\mathrm{opt} \approx \frac{g^2\kappa}{(\omega_m(B_0)-\Delta)^2+\kappa^2}
$.
With this simplified optical diffusion we can find the steady-state variance using Eq.~(\ref{eq: W_{B,SS}}) and  obtain the approximate Fano factor 
\begin{equation}
\label{F1}
F = \frac{(\bar{n}+1)}2 \left(1 -\frac1C+\frac{\Delta-\omega_m+K} \kappa \sqrt{\frac1C-\frac1{C^2}}\right)^{-1}\:,
\end{equation}
which for large cooperativity $C\gg1$ is limited by $F \leq \frac{(\bar{n}+1)}2$.

We find that squeezed number states can be achieved for bath occupation $\bar n <1$. Such small temperatures can be achieved with cryogenic cooling for high mechanical frequencies, but also optomechanical sideband cooling via radiation pressure. This motivates to investigate the case where the oscillator is driven by two lasers, one blue-detuned like here and one red-detuned for additional cooling, in the following Section~\ref{sec: Two independent Cavities}.

\section{Two cavities}

\begin{figure}[t]
\includegraphics[width=0.5\textwidth]{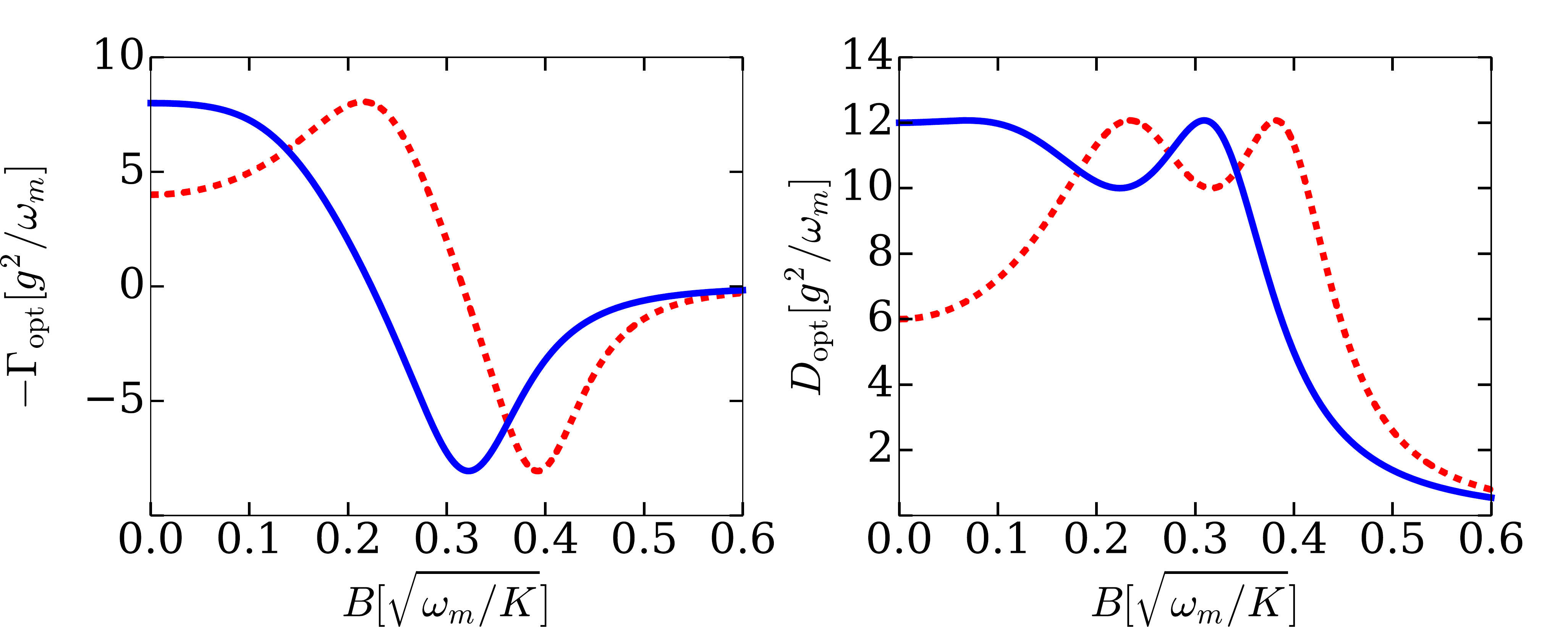}
\caption{\label{Two_cavities}
Optically induced damping $\Gamma_{\mathrm{opt}}$ and diffusion $D_{\mathrm{opt}}$ as a function of the amplitude $B$ for the setup with two lasers fulfilling $\Delta_2 +\Delta_1 =- 2\kappa$. As in Fig. \ref{OneLaserDD}, the off-resonant terms have been neglected. 
 The parameters in this plot are $\kappa/\omega_m =0.1$,  $K/\omega_m \to 0$,  $\Delta_1=\omega_m$ (blue solid line),  $\Delta_1=1.1 \omega_m$ (red dashed line).  
}
\end{figure}

\label{sec: Two independent Cavities}

We now consider the setup depicted in Fig.~ \ref{SetupFig} (b), where a mechanical mode coupled to two cavity modes
$a_1, a_2$. For simplicity, we will here assume these modes to be in separate cavities 
and discuss the corrections arising for the case of a single cavity
driven by two independent lasers in Section \ref{Impl}.  The laser
drive in the first cavity is assumed to be blue-detuned and the laser
in the second cavity red-detuned, i.e. $\Delta_1>0>\Delta_2$. The
second cavity will then induce (positive) optical damping and the
first cavity anti-damping.
The adiabatic elimination is done in analogy to the procedure above
and one finds that both optically induced drift and diffusion are
given by the sum of the individual contributions from Eqs.~\eqref{eq:
  Gamma_opt} and \eqref{DOPT},
i.e.,
$\Gamma_\mathrm{opt}=\Gamma_{\mathrm{opt}}(\Delta_1,
B)+\Gamma_{\mathrm{opt}}(\Delta_2, B)$ and
$D_\mathrm{opt}=D_{\mathrm{opt}}(\Delta_1,
B)+D_{\mathrm{opt}}(\Delta_2, B)$. For simplicity, we assume here
identical $g$ and $\kappa$ in both cavities. The resulting damping and
diffusion are illustrated in
Figure \ref{Two_cavities}.

We are interested in the limit cycle where the average amplitude $B_0$ is the stable solution to $A_B(B_0) = -(\Gamma_m +\Gamma_\mathrm{opt}(B_0)) B_0 = 0$. We approximate the optically induced damping by dropping the non-resonant terms, i.e.
$
\Gamma_{\mathrm{opt}}(B_0) \approx g^2\kappa \left(\frac{1}{(\omega_m(B_0)+\Delta_2)^2+ \kappa^2} -\frac{1}{(\omega_m(B_0)-\Delta_1)^2+ \kappa^2}\right)
$.
Assuming a large cooperativity $C \gg 1$, we can neglect the
mechanical damping and obtain the average amplitude
\begin{equation}
\label{B0eq}
B_0=\frac{1}{2} \sqrt{\frac{\Delta_1-\Delta_2-2\omega_m+2K}{K}}\:,
\end{equation}
valid for $\Delta_1-\Delta_2-2\omega_m+2K>0$. The attractor at $B_0$ is only stable if $|\Delta_1| < |\Delta_2|$ and therefore only then a limit cycle will form. In the following we assume this condition to be satisfied. 

In analogy to the last section, we approximate the steady-state
solution of the amplitude distribution to be a Gaussian centered at
$B_0$. Assuming a large thermal cooperativity, we neglect the mechanically induced diffusion term $\propto \Gamma_m(2\bar{n}+1)$. We also drop the non-resonant terms in the optically induced diffusion so that 
$
D_\mathrm{opt}(B_0)  ={g^2\kappa} \left(\frac{1}{(\omega_m(B_0)-\Delta_1)^2+ \kappa^2}+\frac{1}{(\omega_m(B_0)+\Delta_2)^2+ \kappa^2}\right) 
$.
In  coordinates
$
\Delta_+= \Delta_1 + \Delta_2$ and $\Delta_-=\Delta_1-\Delta_2-2\omega_m+2K
$,
the Fano factor is then given by
\begin{equation}
\begin{split}
F= \frac{D_\mathrm{opt}(B_0)}{B_0 \at{\frac{d\Gamma_{\mathrm{opt}}}{dB}}{B=B_0}}
= -\frac{1}{4}\frac{\Delta_+^2+4\kappa^2}{\Delta_+ \Delta_-}\:.
\end{split}
\end{equation}
We optimize the detunings to achieve a minimal Fano factor:
With respect to $\Delta_+$ it is minimal at $\frac{\mathrm{d}\sigma^2}{\mathrm{d}\Delta_+} = 0$, resulting in
$
\Delta_1 + \Delta_2 =\pm 2\kappa
$.
Note that $\Delta_-$ is always positive since this is the condition to
find the attractor $B_0$. Thus, $\Delta_+$ must be negative, otherwise we would get a negative variance. We therefore choose the solution with the negative sign, i.e., $\Delta_1 + \Delta_2 =- 2\kappa.$ We find the minimal Fano Factor with respect to $\Delta_+$
as 
\begin{equation}
\label{F2}
F  = \frac{\kappa}{2(\Delta_1-\omega_m+\kappa+K)}\:.
\end{equation}
Subpoissonian states ($F<1$) are achieved for a wide set of parameters. In particular for $\Delta_1 \geq \omega_m$ we find non-classical states for any value of $\kappa$, as long as the sidebands are resolved.
In case of small nonlinearities $K  \ll \kappa$ the Fano factor is independent of $K$ in both \eqref{F1} and \eqref{F2}.  

For the system with one laser in Eq.~(\ref{F1})  we found $F=\frac{1}{2}$ for
zero temperature and large cooperativity. In the system with two
lasers we can
achieve even smaller Fano factors by increasing the detuning, but note
that the self-oscillation will not start for too large $\Delta_1$. In both systems the steady state amplitude scales as $B_0^2 \propto 1/K$.

\section{Discussion}
\label{Disc}

\subsection{Comparison to Numerical Results}

\begin{figure}[t]
\includegraphics[width=0.5\textwidth]{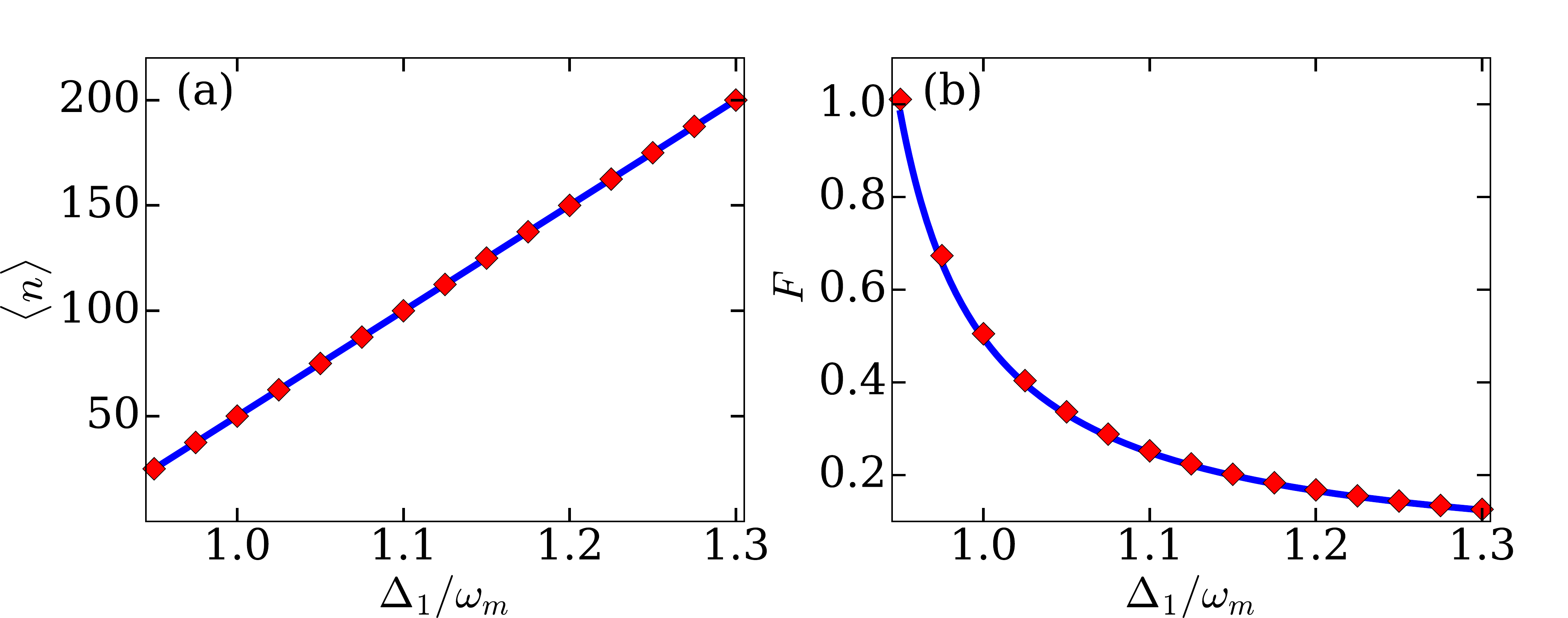}
\caption{ \label{NumFig}
Comparison of analytical approximation from Eqs.~ \eqref{B0eq} and \eqref{F2} 
 to numerical results of (a) mean $\langle n \rangle = \langle b^\dagger b \rangle$  and (b) Fano factor $F=\Delta^2 n/\langle n \rangle$ of the mechanical oscillator in steady state for the setup with two lasers. Here the second laser is tuned to the optimal value $\Delta_2 =-\Delta_1- 2\kappa$ to achieve a small Fano factor. The other parameters in this plot are $n_\mathrm{th}=0$ and
$(K, g, \kappa,
\Gamma_m)/\omega_m=(0.001, 0.001, 0.1, 0)$.
}
\end{figure}

In Fig.~{\ref{NumFig}} we compare our analytical findings of the mean
squared amplitude and variance with an exact numerical solution of the quantum
master equation for the setup with two cavities. We used three states
for each cavity and 60 states for the mechanical oscillator in the
steady-state solver of QuTiP \cite{Johansson2011a, Johansson2013}. The
results are plotted as a function of the cavity decay rate
${\kappa}/{\omega_m}$ and detuning of the first laser
${\Delta_1}/{\omega_m}$. The detuning of the second laser is chosen at
the optimal value $\Delta_2 = -\Delta_1 -2 \kappa$. We assumed a
high-$Q$ oscillator, so that the
intrinsic damping is weak compared to the optically induced damping.
The effective optomechanical coupling was chosen as
${g}/{\omega_m}=0.001$ so that the condition for adiabatic elimination
is approximately fulfilled.  While we are interested here in the limit
of small Kerr nonlinearities, we choose ${K}/{\omega_m}=0.001$ not too
small to keep the Hilbert space small, cf.~Eq.~(\ref{B0eq}). Our
analytical expression leads to values for the Fano factor that are in
excellent agreement with the numerical results.

\subsection{Possible Implementations}

\label{Impl}

In this article we described an ideal system, which yields the simplest analytical description. Depending on the concrete experimental implementation, one has to take into account further corrections, which we discuss in this section.

\textit{Intrinsic Nonlinearity.-}
We assumed a nonlinearity of Kerr type yielding a term $K (b^\dagger b)^2$ in the Hamiltonian. For mechanical oscillators, including also trapped-ion potentials, this stems from a Duffing potential $D (b+b^\dagger)^4$ after a rotating-wave approximation.
 The rotating-wave approximation is valid for $K \ll \omega_m$ and $2K \braket{b^\dagger b} \ll \omega_m$, where $K=6D$. 
 For the system in Section~\ref{OneCav}, i.e. a mechanical oscillator coupled to a driven cavity, we find the condition in the steady state
$
2K \braket{b^\dagger b} \approx  \Delta - \omega_m + K + \kappa \sqrt{C-1} \ll \omega_m.
$
For the two-laser system with optimal detuning relation $\Delta_1 + \Delta_2 =- 2\kappa$ we analogously derive the condition 
$
\Delta_1 -\omega_m+ K  +  \kappa \ll \omega_m
$.
Note that we found in Eqs.~\eqref{F1} and \eqref{F2} the smallest Fano factor for large detuning $\Delta$ or $\Delta_1$, but the rotating-wave approximation is only valid for detuning not much larger than $\omega_m$.  Outside the regime of validity for the rotating wave approximation the Fano factor will significantly larger for a Duffing oscillator than expected from the Kerr-approximation.

\textit{Single Cavity.-}
Instead of driving two separate cavities as proposed in Section \ref{sec: Two independent Cavities}, it may be experimentally simpler to drive a single cavity with two laser tones. The beat between these two frequencies results in additional drift and diffusion terms, which interestingly are phase-dependent. 
The new terms are of the same magnitude as the terms stemming from the
individual lasers, but they rotate at a frequency $\delta=\Delta_1 -
\Delta_2 -2\omega_m(B)$. Therefore we may neglect these terms in a
rotating-wave approximation if $\delta$ is much larger than the optically induced damping and diffusion, corresponding to $\delta \gg g^2/\kappa$. 
On the other hand, tuning $\delta=0$ can lead to rich dynamics in the limit cycle such as a phase-dependent diffusion and damping and therefore phase-dependent squeezing.

\textit{Excitation via Two-Level-Systems.-}
Self-oscillators can also be driven by a two-level system instead of a bosonic mode. In the Hamiltonian this corresponds to a replacement of the annihilation operator $a$ by the Pauli lowering operator $\sigma_-$.
For example, mechanical oscillations of ions can be excited via a
cycling transition \cite{Vahala2009}.
  The results presented above can be transferred to this situation: As we used in the adiabatic elimination only the lowest-order terms in perturbation theory $\propto g^2$, this model is restricted to the lowest two Fock levels of the fast-decaying mode $a$. Therefore adiabatically eliminating a two-level system yields \emph{identical} analytical results.

\textit{Anharmonic Mechanical Oscillators.-}
The most important property of our proposal is the intrinsic mechanical Duffing nonlinearity.
Such nonlinearities can be engineered for example in oscillators made from graphene and carbon nanotubes \cite{Eichler2011, Singh2014a}.
Coupling the oscillator to an auxiliary highly nonlinear system, there have been several proposals to achieve extremely large mechanical nonlinearities  \cite{Jacobs2009,Rips2014, Lu2015}, even on the order of $K/ \omega_m \approx 0.01-0.1$. On this frontier a Duffing nonlinearity tunable by a SQUID was demonstrated in a recent experiment \cite{Ella2015}.
For the motion of trapped ions such high
anharmonicities
of the trapping potential can already be achieved with current systems \cite{Home2011a}.

\subsection{Conclusion and Outlook}

We derived a semiclassical analytical model to characterize self-oscillations for the standard linear optomechanical system with an additional anharmonicity of the mechanical potential. We find excellent agreement with numerical simulations of the system.
The main result is the prediction of a Fano factor
$
F  = \tfrac 12 {\kappa}/{(\Delta_1-\omega_m+\kappa+K)}
$
for a setup using two laser tones at detunings $\Delta_1\approx\omega_m$ and $\Delta_2 =-\Delta_1- 2\kappa$ in the vicinity of the sidebands. For such parameters the Fano factor is nonclassical.

The model derived here can be generalized to other self-oscillators with Duffing nonlinearity such as trapped-ion systems or superconducting circuits. While we have focused  on the amplitude and in particular its steady-state distribution, it will be interesting to describe the phase dynamics in future studies to examine synchronization in the quantum regime.

\section{Acknowledgments}
We would like to acknowledge helpful discussions with E. Amitai,
G. Hegi, A. Mokhberi, A. Nunnenkamp, S. Willitsch, and H. Zoubi. This
work was financially supported by the Swiss SNF and the NCCR Quantum
Science and Technology.

\bibliography{Masterarbeit}

\end{document}